\documentclass[sn-mathphys]{sn-jnl}

\jyear{2021}%

\theoremstyle{thmstyleone}%
\theoremstyle{thmstyletwo}%

\theoremstyle{thmstylethree}%

  \newcommand {\nc} {\newcommand}
  \nc {\beq} {\begin{eqnarray}}
  \nc {\eeq} {\nonumber \end{eqnarray}}
  \nc {\eeqn}[1] {\label {#1} \end{eqnarray}}
  \nc {\eol} {\nonumber \\}
  \nc {\eoln}[1] {\label {#1} \\}
  \nc {\ve} [1] {\mbox{\boldmath $#1$}}
  \nc {\ves} [1] {\mbox{\boldmath ${\scriptstyle #1}$}}
  \nc {\mrm} [1] {\mathrm{#1}}
  \nc {\half} {\mbox{$\frac{1}{2}$}}
  \nc {\thal} {\mbox{$\frac{3}{2}$}}
  \nc {\fial} {\mbox{$\frac{5}{2}$}}
  \nc {\la} {\mbox{$\langle$}}
  \nc {\ra} {\mbox{$\rangle$}}
  \nc {\etal} {\emph{et al.}}
  \nc {\eq} [1] {(\ref{#1})}
  \nc {\Eq} [1] {Eq.~(\ref{#1})}
  \nc {\Refc} [2] {Refs.~\cite[#1]{#2}}
  \nc {\Sec} [1] {Sec.~\ref{#1}}
  \nc {\chap} [1] {Chapter~\ref{#1}}
  \nc {\anx} [1] {Appendix~\ref{#1}}
  \nc {\tbl} [1] {Table~\ref{#1}}
  \nc {\Fig} [1] {Fig.~\ref{#1}}
  \nc {\ex} [1] {$^{#1}$}
  \nc {\Sch} {Schr\"odinger }
  \nc {\flim} [2] {\mathop{\longrightarrow}\limits_{{#1}\rightarrow{#2}}}
  \nc {\IR} [1]{\textcolor{red}{#1}}
  \nc {\IB} [1]{\textcolor{blue}{#1}}
  \nc{\IG}[1]{\textcolor{green}{#1}}
  \nc{\pderiv}[2]{\cfrac{\partial #1}{\partial #2}}
  \nc{\deriv}[2]{\cfrac{d#1}{d#2}}
\nc{\lsim}{\mathrel{\rlap{\lower4pt\hbox{\hskip1pt$\sim$}}
    \raise1pt\hbox{$<$}}} 
  \nc {\elf} {11}
  \nc {\twentytwo} {22}

\begin{document}

\title[Effective field theory analysis of the Coulomb breakup of the one-neutron halo nucleus $^{19}$C]{Effective field theory analysis of the Coulomb breakup of the one-neutron halo nucleus $^{19}$C}


\author*[1]{\fnm{Pierre} \sur{Capel}}\email{pcapel@uni-mainz.de}
\affil[1]{\orgdiv{Institut f\"ur Kernphysik}, \orgname{Johannes Gutenberg-Universit\"at Mainz}, \postcode{55099}, \orgaddress{\city{Mainz}, \country{Germany}}}

\author*[2]{\fnm{Daniel R.} \sur{Phillips}}\email{phillid1@ohio.edu}
\affil[2]{\orgdiv{Institute of Nuclear and Particle Physics and Department of Physics and Astronomy}, \orgname{Ohio University}, \orgaddress{\city{Athens}, \postcode{45701}, \state{OH}, \country{USA}}}

\author[3]{\fnm{Andrew} \sur{Andis}}
\affil[3]{\orgdiv{Department of Physics}, \orgname{North Carolina State University}, \orgaddress{\city{Raleigh}, \postcode{27695-8202}, \state{NC}, \country{USA}}}

\author[4]{\fnm{Mirko} \sur{Bagnarol}}
\affil[4]{\orgdiv{Racah Institute of Physics}, \orgname{Hebrew University Of Jerusalem}, \orgaddress{\city{Jerusalem}, \postcode{9190401}, \country{Israel}}}

\author[5]{\fnm{Behnaz} \sur{Behzadmoghaddam}}
\affil[5]{\orgdiv{Department of Physics}, \orgname{University of Tehran}, \postcode{P.O. Box 14395-547}, \orgaddress{\city{Tehran}, \country{Iran}}}

\author[1,6]{\fnm{Francesca} \sur{Bonaiti}}
\affil[6]{\orgdiv{PRISMA+ Cluster of Excellence}, \orgname{Johannes Gutenberg-Universit\"at Mainz}, \postcode{55099}, \orgaddress{\city{Mainz}, \country{Germany}}}

\author[7]{\fnm{Rishabh} \sur{Bubna}	}
\affil[7]{\orgdiv{Helmholtz-Institut f\"ur Strahlen- und Kernphysik}, \orgname{Rhenish Friedrich Wilhelm University}, \postcode{53115}, \orgaddress{\city{Bonn}, \country{Germany}}}

\author[8,9]{\fnm{Ylenia} \sur{Capitani}}
\affil[8]{\orgdiv{Dipartimento di Fisica}, \orgname{Universit\`a di Trento}, \postcode{38123}, \orgaddress{\city{Trento}, \country{Italy}}}
\affil[9]{\orgdiv{INFN-TIFPA Trento Institute for Fundamental Physics and Applications}, \postcode{38123}, \orgaddress{\city{Trento}, \country{Italy}}}

\author[10,\elf]{\fnm{Pierre-Yves} \sur{Duerinck}}
\affil[10]{\orgdiv{Physique Nucl\'eaire et Physique Quantique (CP 229)}, \orgname{Universit\'e libre de Bruxelles}, \postcode{1050}, \orgaddress{\city{Brussels}, \country{Belgium}}}
\affil[\elf]{\orgdiv{Institut Pluridisciplinaire Hubert Curien (IPHC), CNRS/IN2P3}, \orgname{Universit\'e de Strasbourg}, \postcode{67037}, \orgaddress{\city{Strasbourg}, \country{France}}}

\author[1]{\fnm{Victoria} \sur{Durant}}

\author[12]{\fnm{Niklas} \sur{Döpper}}
\affil[12]{\orgdiv{Physics Department}, \orgname{Technische Universität München}, \postcode{85748}, \orgaddress{\city{Garching}, \country{Germany}}}

\author[13]{\fnm{Aya} \sur{El Boustani}}
\affil[13]{\orgdiv{Departamento de Física Atómica, Molecular y Nuclear}, \orgname{Universidad de Sevilla}, \postcode{41012}, \orgaddress{\city{Seville}, \country{Spain}}}

\author[14]{\fnm{Roland} \sur{Farrell}}
\affil[14]{\orgdiv{Department of Physics}, \orgname{University of Washington}, \orgaddress{\city{Seattle}, \postcode{98195-1560}, \state{WA}, \country{USA}}}

\author[12]{\fnm{Maurus} \sur{Geiger}}

\author[15,16]{\fnm{Michael} \sur{Gennari}}
\affil[15]{\orgdiv{Department of Physics and Astronomy}, \orgname{University of Victoria}, \orgaddress{\city{Victoria}, \postcode{V8P\ 5C2}, \state{BC}, \country{Canada}}}
\affil[16]{\orgname{TRIUMF}, \orgaddress{\city{Vancouver}, \postcode{V6T 2A3}, \state{BC}, \country{Canada}}}

\author[4]{\fnm{Nitzan} \sur{Goldberg}}

\author[17]{\fnm{Jakub} \sur{Herko}}
\affil[17]{\orgdiv{Department of Physics and Astronomy}, \orgname{University of Notre Dame}, \orgaddress{\city{Notre Dame}, \postcode{46556}, \state{IN}, \country{USA}}}

\author[18]{\fnm{Tanja} \sur{Kirchner}}
\affil[18]{\orgdiv{Institut f\"ur Kernphysik}, \orgname{Technische Universit\"at Darmstadt}, \postcode{64289}, \orgaddress{\city{Darmstadt}, \country{Germany}}}

\author[1]{\fnm{Live-Palm} \sur{Kubushishi}}

\author[19]{\fnm{Zhen} \sur{Li}}
\affil[19]{\orgdiv{Laboratoire de Physique des Deux Infinis Bordeaux (LP2I Bordeaux)}, \orgname{Universit\'e de Bordeaux, CNRS/IN2P3}, \postcode{33175}, \orgaddress{\city{Gradignan cedex}, \country{France}}}

\author[1]{\fnm{Simone S.} \sur{Li Muli}}

\author[20]{\fnm{Alexander} \sur{Long}}
\affil[20]{\orgdiv{Department of Physics}, \orgname{The George Washington University}, \orgaddress{\city{Washington}, \postcode{20052}, \state{DC}, \country{USA}}}

\author[21]{\fnm{Brady} \sur{Martin}}
\affil[21]{\orgdiv{Department of Physics and Astronomy}, \orgname{University of Iowa}, \orgaddress{\city{Iowa City}, \postcode{52242}, \state{IA}, \country{USA}}}

\author[\twentytwo]{\fnm{Kamyar} \sur{Mohseni}}
\affil[\twentytwo]{\orgdiv{Department of Physics}, \orgname{Aeronautics Institute of Technology (ITA), DCTA}, \orgaddress{\city{S\~ao Jos\'e dos Campos}, \postcode{12228-900}, \state{SP}, \country{Brazil}}}

\author[23]{\fnm{Imane} \sur{Moumene}}
\affil[23]{\orgdiv{Galileo Galilei Institute}, \orgname{Istituto Nazionale di Fisica Nucleare (INFN)}, \postcode{50125}, \orgaddress{\city{Florence}, \country{Italy}}}

\author[8]{\fnm{Nicola} \sur{Paracone}}

\author[4]{\fnm{Elad} \sur{Parnes}}

\author[24]{\fnm{Beatriz} \sur{Romeo}}
\affil[24]{\orgname{Donostia International Physics Center (DIPC)}, \postcode{20018}, \orgaddress{\city{Donostia-San Sebasti\`an}, \country{Spain}}}
     
\author[25]{\fnm{Victor} \sur{Springer}}
\affil[25]{\orgdiv{Institut f\"ur Theoretische Physik II}, \orgname{Ruhr-Universit\"at}, \postcode{44780}, \orgaddress{\city{Bochum}, \country{Germany}}}

\author[26]{\fnm{Isak} \sur{Svensson}}
\affil[26]{\orgdiv{Department of Physics}, \orgname{Chalmers University of Technology}, \postcode{412~96}, \orgaddress{\city{Gothenburg}, \country{Sweden}}}

\author[26]{\fnm{Oliver} \sur{Thim}}

\author[3]{\fnm{Nuwan} \sur{Yapa}}


\abstract{We analyse the Coulomb breakup of $^{19}$C measured at $67A$ MeV at RIKEN.
We use the Coulomb-Corrected Eikonal (CCE) approximation to model the reaction and describe the one-neutron halo nucleus $^{19}$C within Halo Effective Field Theory (EFT).
At leading order we obtain a fair reproduction of the measured cross section as a function of energy and angle. The description is insensitive to the choice of optical potential, as long as it accurately represents the size of ${}^{18}$C. It is also insensitive to the interior of the ${}^{19}$C wave function. Comparison between theory and experiment thus enables us to infer asymptotic properties of the ground state of $^{19}$C: these data put constraints on the one-neutron separation energy of this nucleus and, for a given binding energy, can be used to extract an asymptotic normalisation coefficient (ANC). 
These results are confirmed by CCE calculations employing next-to-leading order Halo EFT descriptions of ${}^{19}$C: at this order the results for the Coulomb breakup cross section are completely insensitive to the choice of the regulator.
Accordingly, this reaction can be used to constrain the one-neutron separation energy and ANC of ${}^{19}$C.}

\keywords{Halo nuclei, Coulomb breakup, Halo Effective Field Theory, Eikonal approximation, $^{19}$C}



\maketitle

\section{Introduction}\label{intro}

Measurements of nuclear reactions along several isotopic chains show that the neutron distribution becomes extended as the neutron dripline is approached \cite{Tan85b,Tan85l}. This has led to the identification of ``neutron halos": situations where a significant fraction of the neutron probability distribution resides in the classically forbidden region \cite{Tan96}. Up to $Z=6$ we already have examples of four-neutron halos, e.g., ${}^8$He, two-neutron halos, e.g., ${}^{22}$C, ${}^{19}$B, and one-neutron halos, e.g., ${}^{11}$Be, ${}^{19}$C. 

This last nucleus demonstrates the striking features of a one-neutron halo.
Following the dissociation of the halo neutron from the $^{18}$C core, the momentum distribution of either of these fragments 
is narrow \cite{Baz95,Mar96,BBB98,Hwa17}, as one would expect from a spatially extended system.
Moreover, the breakup cross section of this fragile structure is large \cite{Nakamura:1999rp,Nakamura:2003cyk,Sat08}.
This is particularly true on a heavy target such as Pb, for which the reaction is strongly Coulomb dominated.
In that case an enhanced E1 strength is observed at low core-neutron relative energy, which is sometimes called the ``pygmy dipole resonance".  
It is perhaps counter-intuitive that properties of the neutron distribution can be probed through an electromagnetic observable, but this significant low-energy E1 strength is a consequence of the extended neutron distribution dragging the center-of-mass of the halo away from the center-of-charge. It is thus related to the halo physics that yields a significant isotope shift in these systems---and this relation can be formalised through the non-energy-weighted sum rule. For $s$-wave one-neutron halos this  physics is ``universal" in the sense that it depends only on the one-neutron separation energy and the Asymptotic Normalisation Coefficient (ANC) of the ground-state wave function.

The Coulomb dissociation of ${}^{19}$C was measured by Nakamura {\it et al.} at RIKEN at $67A$ MeV already in the last millenium~\cite{Nakamura:1999rp,Nakamura:2003cyk}. The large E1 strength below 1 MeV outgoing relative energy of the ${}^{18}$C-neutron system indicates the presence of a neutron halo. Comparison with models of the reaction implied that this is an $s$-wave halo, with a one-neutron separation energy $S_{\rm n}=530 \pm 130$ keV.

Halo Effective Field Theory (Halo EFT) provides a systematic way to analyse the Coulomb dissociation of one-neutron halos. (For a general introduction to Halo EFT and a review of the method's status as it stood in 2017, see Ref.~\cite{Hammer:2017tjm}.) Halo EFT expands the amplitude for the nuclear reaction in powers of the expansion parameter $R_{\rm core}/R_{\rm halo}$, where, in this case, $R_{\rm core}$ is the size of ${}^{18}$C, which amounts to approximately $2.5$ fm, and $R_{\rm halo}$ is the size of the neutron halo in ${}^{19}$C, estimated to be about 6.5 fm. The calculation of Coulomb dissociation in Halo EFT confirms that the amplitude is universal at leading order, depending only on $S_{\rm n}$ and the charge-to-mass ratio of the target~\cite{Hammer:2011ye,Acharya:2013nia,Hammer:2017tjm}. At next-to-leading order the asymptotic normalisation coefficient of the halo affects the amplitude. But, once $S_{\rm n}$ and the ANC are fixed the amplitude is predicted---at least for $s$-wave halos---up to errors of order $\left(\frac{R_{\rm core}}{R_{\rm halo}}\right)^3$ in the Halo EFT expansion. In Ref.~\cite{Acharya:2013nia} the photodissociation of ${}^{19}$C was computed in Halo EFT and the equivalent photon approximation was used to convert the photodissociation cross section of ${}^{19}$C into a Coulomb-breakup cross section. Acharya and Phillips extracted the value $S_{\rm n}=575 \pm 55 {\rm (stat.)} \pm 20 {\rm (EFT)}$\,MeV from the low-energy ($E < 1$ MeV) and small-angle ($\theta < 2.2^\circ$) portion of the data from Ref.~\cite{Nakamura:1999rp}. 

The reaction-theory employed in Ref.~\cite{Acharya:2013nia} was quite rudimentary. In this work we couple an EFT description of the ${}^{19}$C bound state to a more advanced treatment of the reaction on the ${}^{208}$Pb target
that uses the Coulomb-Corrected Eikonal approximation (CCE) \cite{MBB03,CBS08,BC12}.
This approximation corrects the erroneous treatment of the Coulomb interaction within the usual eikonal description of breakup reactions.
It enables a computation of breakup cross sections at intermediate beam energies on both light and heavy targets that attains excellent agreement with fully dynamical reaction models while also retaining the simplicity and numerical efficiency of the usual eikonal approximation \cite{CBS08}.
Within this implementation of the CCE, the ${}^{19}$C bound state, and the ${}^{18}$C-neutron continuum, are described within Halo EFT, viz.\ using a set of ${}^{18}$C-neutron potentials of Gaussian shape. In a leading-order calculation the depth of the Gaussian is adjusted to reproduce a particular $S_{\rm n}$. In a next-to-leading-order calculation, an additional term is added to the potential, and its parameter is adjusted to produce a specific ANC. Performing the calculation for a range of Gaussian widths checks whether the breakup cross section is insensitive to details of the potential. This imitates the strategy successfully employed for ${}^{11}$Be and ${}^{15}$C reactions on various targets in Refs.~\cite{CPH18,YC18,MYC19,HC21}. 
Coupling a reliable model of the reaction to a Halo-EFT description of the nucleus provides a detailed account of the reaction mechanism, while enabling us to study very systematically the influence that the halo nucleus' structure has on the reaction cross sections.

The calculations described in this paper were initially performed as part of a week-long set of exercises at the TALENT school ``Effective Field Theories in Light Nuclei: from Structure to Reactions'' that took place at the Mainz Institute for Theoretical Physics in July--August 2022 \cite{TALENT22}. 
Students at the school (the majority of the authors in this paper) tuned the Gaussian potentials to reproduce specific scattering and bound-state parameters for the ${}^{19}$C system. They then ran CCE calculations, predicted the Coulomb dissociation cross sections, and compared the result with data. The following sections describe their work and its outcomes, as follows. In Sec.~\ref{CCE} we provide a brief summary of the reaction model and its implementation within the CCE. Section~\ref{LO} lays out the leading-order (LO) calculation, presenting results for the ${}^{19}$C system for Gaussians of widths ranging from 0.5 to 2.5~fm. These potentials are then used, together with the CCE, to predict the Coulomb breakup cross section. We find that the cross section scales with the square of the ${}^{19}$C ANC, demonstrating that the reaction is almost exclusively peripheral. In Sec.~\ref{optpot} we explore the sensitivity of the results to the optical potentials chosen for the ${}^{18}$C-${}^{208}$Pb and neutron-${}^{208}$Pb systems. In Sec.~\ref{NLO} we confirm that the cross section is insensitive to the interior of the ${}^{18}$C-neutron wave function by performing a NLO calculation and finding (almost) the same result irrespective of the width of the Gaussian employed. In Sec.~\ref{BE} we return to the LO potentials and vary the binding energy, in order to check the confidence interval given by Acharya and Phillips in Ref.~\cite{Acharya:2013nia}. Finally, in Sec.~\ref{Conclusion} we offer some conclusions and point out some interesting aspects of the EFT description of these Coulomb-dissociation data that, we believe, can motivate further theoretical and experimental studies of ${}^{19}$C.

\section{Reaction model}\label{model}
\subsection{Three-body model of Coulomb breakup}
To describe the breakup of $^{19}$C on ${}^{208}$Pb, we consider the usual three-body model of the reaction \cite{BC12}.
The projectile $P$ is seen as a two-body structure: a halo neutron (n of mass $m_{\rm n}$ and charge nil) loosely-bound to a $^{18}$C core assumed to be in its $0^+$ ground state ($c$ of mass $m_c$ and charge $Z_c e$).
This two-body structure is described by the effective Hamiltonian
\beq
H_0=-\frac{\hbar^2}{2\mu}\Delta_r+V_{c\rm n}(r),
\eeqn{e01}
where $\ve{r}$ is the $c$-n relative coordinate, $\mu=m_cm_{\rm n}/(m_c+m_{\rm n})$ is the $c$-n reduced mass, and $V_{c\rm n}$ is an effective potential that describes the $c$-n interaction.
As discussed in Secs.~\ref{LOdescr} and \ref{NLO}, we consider Halo-EFT interactions up to NLO \cite{Bertulani:2002sz,Hammer:2017tjm}.

The eigenstates $\phi$ of $H_0$ describe the different states of the projectile.
The negative-energy eigenstates correspond to the $c$-n bound states.
They are discrete and, in addition to quantum numbers of the $c$-n orbital angular momentum $l$, the total angular momentum $j$, and its projection $m$, they are identified by the number of nodes in their radial wave function $n_r$.
Asymptotically, the radial part of these bound-state wave functions behaves as
\beq
u_{n_rlj}(r)\flim{r}{\infty}{\cal C}_{n_rlj}\ \kappa_{n_rlj}\,r\ k_l(\kappa_{n_rlj}\,r),
\eeqn{e01b}
where $k_l$ is a modified spherical Bessel function of the second kind, $\kappa_{n_rlj}$ is related to the eigenenergy of the state $E_{n_rlj}=-\hbar^2\kappa^2_{n_rlj}/2\mu$, and ${\cal C}_{n_rlj}$ is the asymptotic normalisation coefficient (ANC) associated with that bound state.
The positive-energy eigenstates describe the $c$-n continuum part of the projectile spectrum, viz. the broken up projectile.
As such they are identified by their $c$-n relative energy $E$, in addition to the quantum number defining the partial wave $l$, $j$, and $m$.

The lead target $T$ is assumed to be a structureless cluster of mass $m_T$ and charge $Z_T e$.
Its interaction with the projectile components $c$ and n is described by the optical potentials $V_{cT}$ and $V_{{\rm n} T}$, respectively.
These potentials are found in the literature as explained in \Sec{optpot}.

Within this three-body model of the collision, studying the $P$-$T$ collision reduces to solving the following \Sch equation
\beq
H\ \Psi(\ve{r},\ve{R})= E_T\ \Psi(\ve{r},\ve{R}),
\eeqn{e02}
with the three-body Hamiltonian
\beq
H= -\frac{\hbar^2}{2\mu_{PT}}\Delta_R+H_0+V_{cT}(R_{cT})+V_{{\rm n}T}(R_{{\rm n}T}),
\eeqn{e03}
where $\ve{R}$ is the coordinate of the projectile center of mass relative to the target, $\mu_{PT}=m_Pm_T/(m_P+m_T)$ is the $P$-$T$ reduced mass---with $m_P=m_c+m_{\rm n}$---and $\ve{R}_{cT}$, resp. $\ve{R}_{{\rm n}T}$, are the $c$-$T$, resp. n-$T$, relative coordinates.
The total energy $E_T$ in \Eq{e02} is related to the initial $P$-$T$ kinetic energy and the eigenenergy of the projectile in its initial ground state $\phi_{n_{r0}l_0j_0m_0}$ through
\beq
E_T=\frac{\hbar^2}{2\mu_{PT}}K_0^2+E_{n_{r0}l_0j_0},
\eeqn{e04}
where $\ve{K}_0$ is the wave vector of the incoming $P$-$T$ relative motion; that direction defines the $Z$ axis of the system of coordinates.

The \Sch equation \eq{e02} has to be solved with the incoming condition that the projectile, in its ground state, is impinging on the target.
Accordingly, the three-body wave function behaves as
\beq
\Psi^{(m_0)}(\ve{r},\ve{R})\flim{Z}{-\infty}e^{iK_0Z}\phi_{n_{r0}l_0j_0m_0}(\ve{r}).
\eeqn{e05}
Various numerical techniques, based on different approximations, have been developed to solve this equation, see Ref.~\cite{BC12} for a recent review.
For this study, we have considered the CCE \cite{MBB03,CBS08,BC12}, which is very efficient at the intermediate beam energy considered here.

\subsection{Coulomb-Corrected Eikonal approximation}\label{CCE}
At sufficiently high energy, the eikonal approximation is quite reliable to describe the $P$-$T$ collision \cite{Glauber,BC12}.
Within that approximation, the three-body wave function after the collision reads
\beq
\Psi^{(m_0)}(\ve{r},\ve{R})\flim{Z}{\infty}e^{iK_0Z}e^{i\chi(\ve{r},\ve{R})}\phi_{n_{r0}l_0j_0m_0}(\ve{r}),
\eeqn{e06}
where the eikonal phase is given by
\beq
\chi(\ve{r},\ve{R})=-\frac{1}{\hbar v}\int_{-\infty}^{\infty}\left[V_{cT}(R_{cT})+V_{{\rm n}T}(R_{{\rm n}T})\right]\,dZ,
\eeqn{e07}
with $v=\hbar K_0/\mu_{PT}$ the $P$-$T$ initial velocity.

Being based on an adiabatic description of the reaction, the usual eikonal approximation is valid only for reactions that take place over a short time, viz. that are dominated by the short-ranged nuclear interaction.
When the Coulomb interaction is non-negligible, such as for the lead target considered in this study, the breakup cross section inferred from the expression \eq{e06} diverges \cite{MBB03,CBS08}.
Margueron \etal\ have developed a correction, that efficiently solves that divergence \cite{MBB03}.
The main idea is to use the first order of the perturbation theory, which accounts for the  projectile dynamics, to correct for the erroneous treatment of the Coulomb interaction at the eikonal approximation.
In the CCE, the Coulomb contribution to the eikonal phase is replaced, at first order, by its corresponding perturbative estimate \cite{CBS08}:
\beq
e^{i\chi(\ve{r},\ve{R})}&\stackrel{\rm CCE}{\longrightarrow}&e^{i\chi_{\rm N}(\ve{r},\ve{R})}\left[e^{i\chi_{\rm C}(\ve{r},\ve{R})}-i\chi_{\rm C}(\ve{r},\ve{R})+i\chi_{\rm FO}(\ve{r},\ve{R})\right],
\eeqn{e08}
where $\chi_{\rm N}$ and $\chi_{\rm C}$ are, respectively, the nuclear and Coulomb contributions to the eikonal phase $\chi$ \eq{e07}, and where the first-order phase reads
\beq
\chi_{\rm FO}(\ve{r},\ve{R})=-\eta\int_{-\infty}^{\infty}e^{i\omega Z/v}\left(\frac{1}{R_{cT}}-\frac{1}{R}\right)\,dZ,
\eeqn{e09}
with $\eta=Z_cZ_T e^2/(4\pi\epsilon_0 \hbar v)$ the $P$-$T$ Sommerfeld parameter and $\hbar \omega=E-E_{n_{r0}l_0j_0}$ the energy difference between the final continuum state and the initial bound state of the projectile.

By accounting for the projectile dynamics in the first-order treatment of the Coulomb interaction, this correction solves the aforementioned divergence issue.
Moreover the expression \eq{e08} enables us to account also for the nuclear part of the $P$-$T$ interaction at all orders, its interference with the Coulomb force, and, although only in an approximate way, for higher-order Coulomb effects.
This CCE leads to breakup cross sections in excellent agreement with fully dynamical models \cite{CBS08}.
It is thus well suited to describe breakup reactions at intermediate energies on both light and heavy targets, while exhibiting the simplicity and numerical efficiency of a usual eikonal code.
In this study we consider the CCE to compute the breakup cross section of $^{19}$C impinging on ${}^{208}$Pb at $67A$~MeV and compare these theoretical results with the data of Ref.~\cite{Nakamura:1999rp}.

\section{Leading-order calculation}\label{LO}

\subsection{Leading-order description of $^{19}$C}\label{LOdescr}

The one-neutron halo nucleus $^{19}$C has a $\half^+$ ground state that lies slightly more than half an MeV below the one-neutron separation threshold ($S_{\rm n}= 0.58\pm0.09$~MeV \cite{Masses212}).
Various experiments have confirmed the one-neutron halo structure of that state \cite{Baz95,Mar96,BBB98,Nakamura:1999rp,Nakamura:2003cyk,Sat08,Hwa17}.
Therefore it is usually described as a $^{18}$C in its $0^+$ ground state to which a valence neutron is loosely bound in the $s_{1/2}$ partial wave.
As mentioned earlier, we consider in the present study a Halo-EFT description of $^{19}$C, assuming the halo neutron sits in a $0s_{1/2}$ bound state, i.e., with $n_r=0$ nodes in the radial wave function.

At leading order the ${}^{18}$C-n interaction is described by a Gaussian potential:
\begin{equation}
V_{\rm LO}(r;\sigma)=C_0(\sigma) \frac{1}{(2 \pi \sigma^2)^{3/2}}\exp\left(-\frac{r^2}{2 \sigma^2}\right),\label{e1}
\end{equation}
where the standard deviation of the Gaussian, $\sigma$, acts as a regulator. In the limit $\sigma \rightarrow 0$ this becomes a three-dimensional $\delta$-function. We consider 
 $\sigma=0.5$, 1, 1.5, 2, 2.5~fm. For each $\sigma$, the potential strength $C_0$ is adjusted to produce a $0s_{1/2}$ ${}^{19}$C state that is bound by 0.58~MeV with respect to the ${}^{18}$C-neutron threshold.
The corresponding values of the $C_0$s and the ANCs ${\cal C}_{0s1/2}$ predicted for ${}^{19}$C are given in Table~\ref{t1}.

\begin{table}
\center
\begin{tabular}{c|cc}
\hline\hline
$\sigma$ & $C_0$ & ${\cal C}_{0s1/2}$ \\
(fm) & (MeV\, fm$^3$) & (fm$^{-1/2}$) \\ \hline
0.5 & -262.25 & 0.620 \\
1 & -590.93 & 0.673 \\
1.5 & -992.67 & 0.731 \\
2 & -1474.16 & 0.793 \\
2.5 & -2042.53 & 0.861 \\ \hline\hline
\end{tabular}
\caption{\label{t1}Strengths of the LO $^{18}$C-n potentials for the different regulators $\sigma$ considered in this study [see \Eq{e1}]. They have been fitted to reproduce the ground state energy at $E_{0s1/2}=-0.58$~MeV. The corresponding ANCs ${\cal C}_{0s1/2}$ are listed as well.}
\end{table}

\begin{figure}[h]%
\centering
\includegraphics[width=0.49\textwidth]{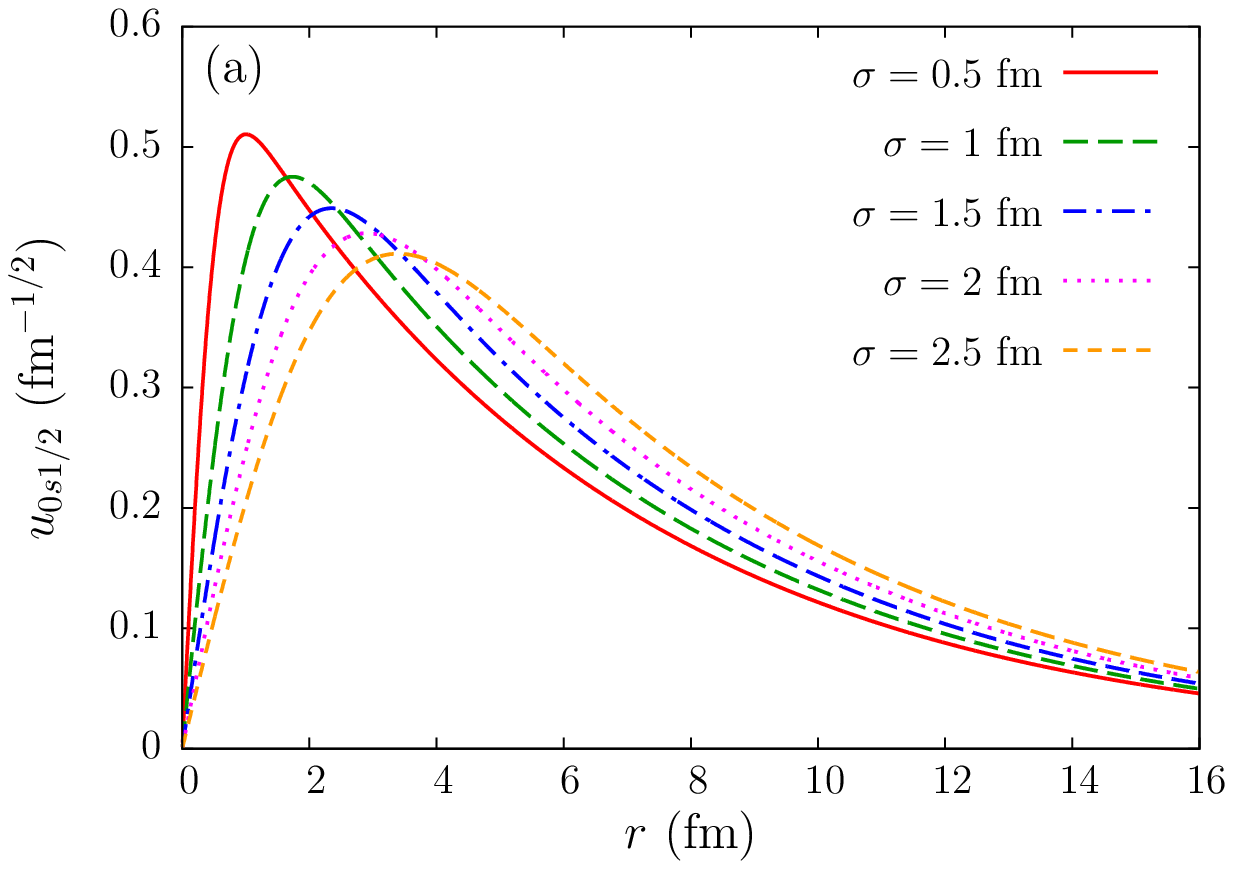}
\includegraphics[width=0.49\textwidth]{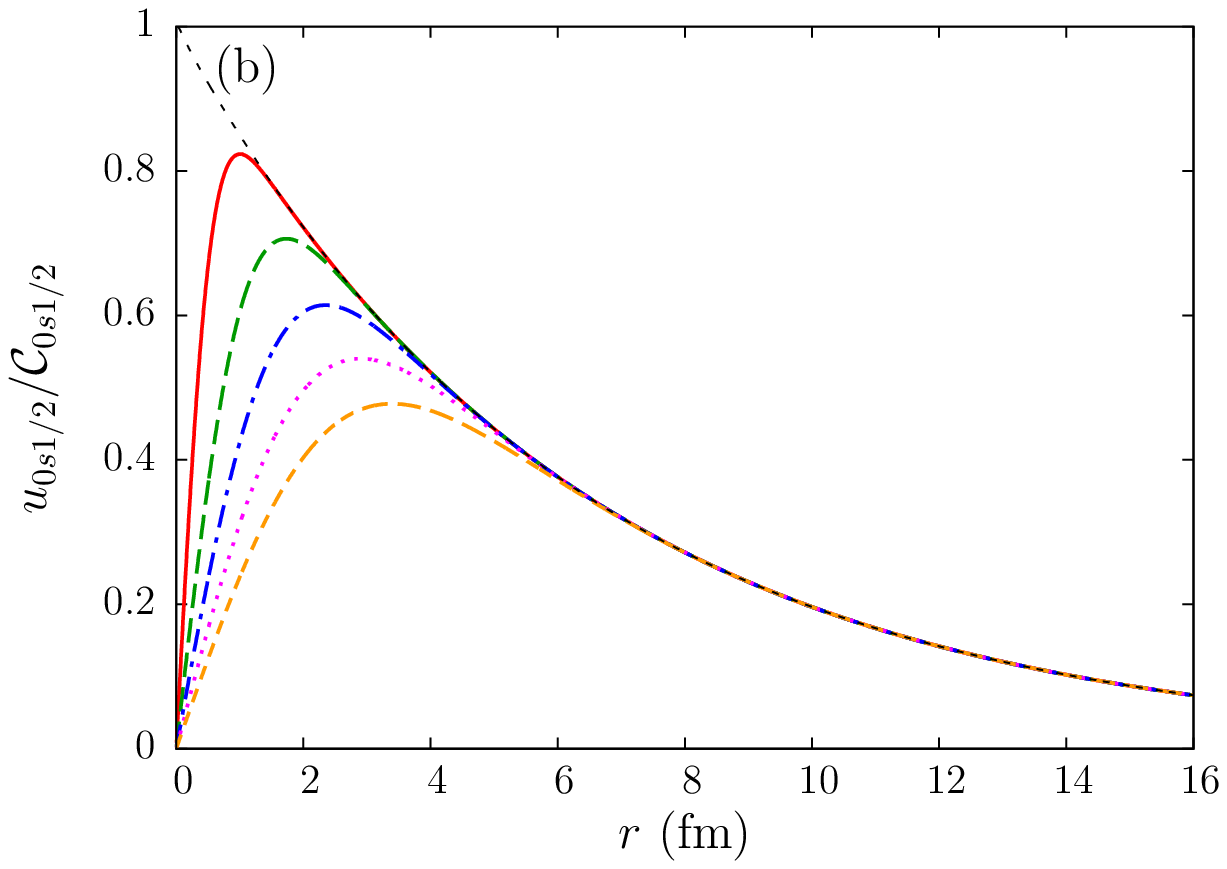}
\caption{Reduced radial wave functions of the $0s_{1/2}$ $^{18}$C-n bound state (a) normalised to unity; (b) divided by their ANC ${\cal C}_{0s1/2}$ for different values of $\sigma$, as indicated in the legend.}\label{f1}
\end{figure}

Figure~\ref{f1} shows the reduced radial wave functions obtained within this LO Halo-EFT model of $^{19}$C 
(a) normalised to unity and (b) divided by their ANC.
Panel (b) shows that the wave functions have the same asymptotic behaviour (thin black dashed line), up to a multiplicative constant---as should be the case given the way in which they were constructed.
It also shows they differ markedly at short range, viz. for $r \lsim 3$~fm. This provides a straightforward way to test if the reaction is peripheral: if it is, the reaction cross section will scale as 
the square of the ANC. If the reaction is sensitive to the short-range piece of the ${}^{18}$C-n wave functions then that scaling will break down.

\subsection{Coulomb breakup cross sections at LO}

\begin{figure}[h]%
\centering
\includegraphics[width=0.49\textwidth]{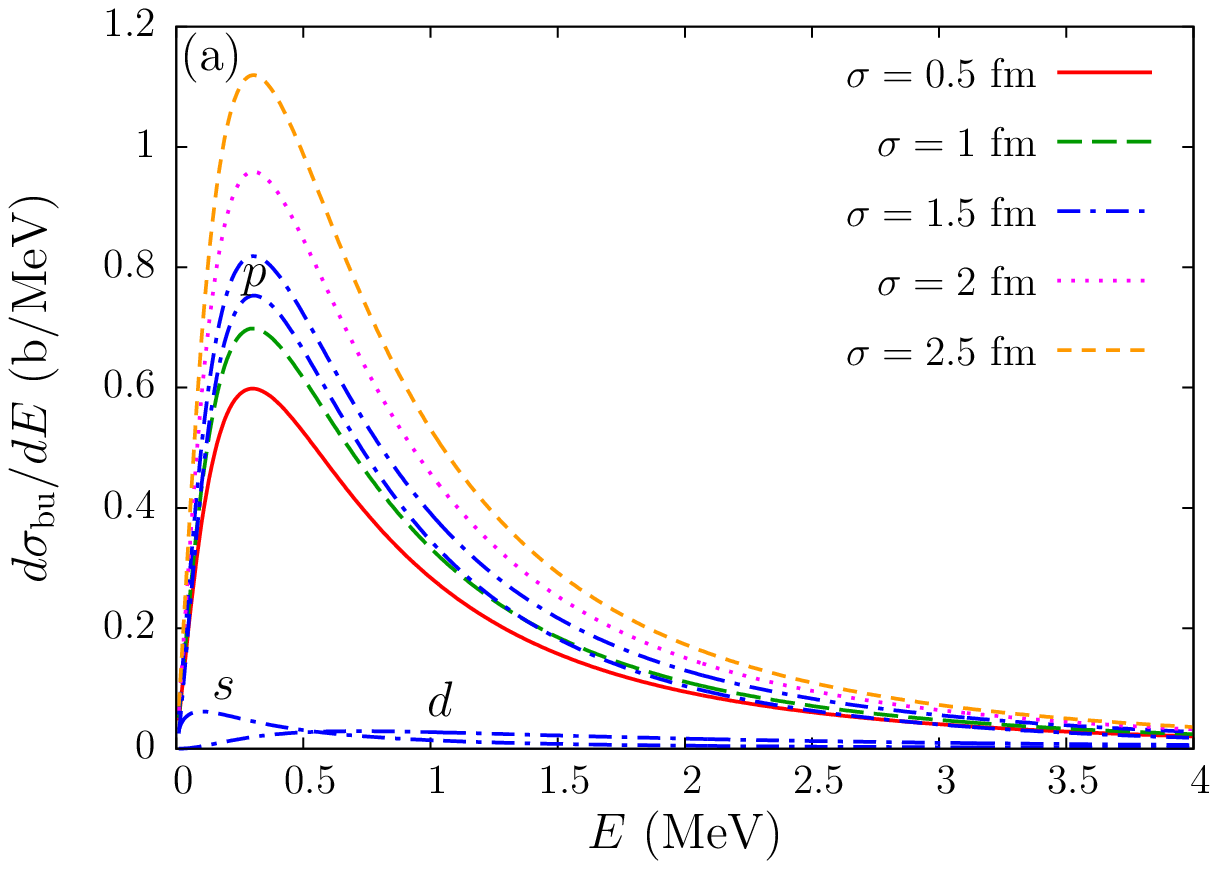}
\includegraphics[width=0.49\textwidth]{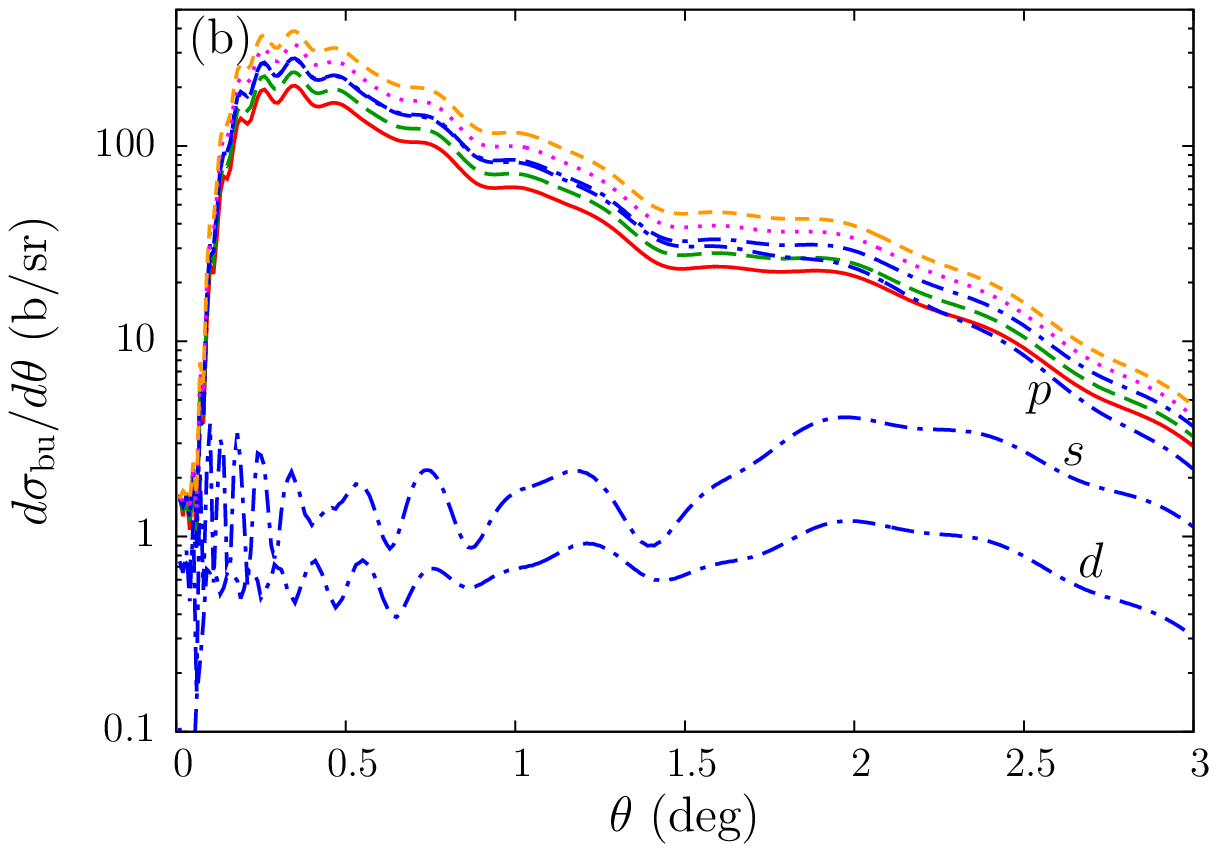}
\caption{Breakup cross section of $^{19}$C on ${}^{208}$Pb at $67A$ MeV
(a) plotted as a function of the $^{18}$C-n relative energy $E$ after dissociation, and (b) plotted as a function of the scattering angle of the $^{18}$C-n centre of mass for energies $0\le E\le0.5\,{\rm MeV}$.
In both cases, $s$, $p$, and $d$ components are shown separately for the $\sigma=1.5$~fm case (blue dash-dotted lines).}\label{f2}
\end{figure}

We now execute the CCE code with the LO $^{18}$C-n potentials of \Sec{LOdescr}\footnote{An open-access version of the {\sc fortran} code ``Chaconne'' has been developed for the TALENT school ``Effective Field Theories in Light Nuclei: from Structure to Reactions'' \cite{TALENT22}.
The program, a user's manual, test input file, and the corresponding output, can be 
downloaded from the school website, see the documents attached to the lecture on nuclear reaction theory in the third week of the school \url{https://indico.mitp.uni-mainz.de/event/279/timetable/\#20220808}.}.
As explained in \Sec{model}, the interactions between the projectile constituents and the target are simulated by optical potentials selected from the literature.
The reasons for this selection, and its effect on our calculations, will be discussed in Sec.~\ref{optpot}. 
Figure~\ref{f2} gives the direct CCE results---viz. without data---for the five values of the Gaussian range $\sigma$ considered in \Sec{LOdescr}.
The breakup cross section plotted as a function of the $^{18}$C-n relative energy after dissociation is shown in \Fig{f2}(a), whereas \Fig{f2}(b) displays it as a function of the scattering angle of the $^{18}$C-n centre of mass for a continuum energy $0\le E\le0.5$~MeV.
For $\sigma=1.5$~fm (blue dash-dotted lines), the contributions to the cross section from $s$, $p$, and $d$ waves in the ${}^{18}$C-n continuum are shown separately. 
It is immediately clear that the reaction is dominated by an E1 transition from the $s$ ground state to the $p$ continuum, as expected for the part of the cross section mediated by a single E1 photon exchange between the
${}^{208}$Pb nucleus and the ${}^{19}$C projectile. For our LO calculation we take the $p$-wave phase shifts to be 0, because there is no known state with negative parity at low energy.
 This makes the overall result rather simple, cf. Eq.~(15) of Ref.~\cite{Acharya:2013nia}.

However, the presence of nuclear interactions between ${}^{18}$C and ${}^{208}$Pb and between the neutron and ${}^{208}$Pb, as well as the possibility of multiple photon exchanges, produce noticeable contributions to the cross section from $s$ and $d$ waves in the ${}^{18}$C-neutron continuum.
Both $s$- and $d$-wave contributions become a larger fraction of the breakup cross section as the angle increases, although the $d$-wave piece stays a factor of a few below the $s$-wave one throughout the angular range of interest here.
The $s$-wave effect is more important at lower relative energy, with the $d$-wave one growing as energy increases.
This is a significant finding, because the CCE is nearly as simple mathematically as the first-order E1 treatment carried out in Ref.~\cite{Acharya:2013nia}, but it allows us to quantify the nuclear contribution to breakup, its interference with the Coulomb force, and other quantal interferences seen in the oscillatory pattern of the angular distribution \cite{CBS08}.

While the way that the reaction mechanism populates different partial waves in the continuum is interesting, the key finding from Fig.~\ref{f2} is that population of anything other than the continuum $p$ wave is small enough that the total cross section scales (nearly) perfectly with ${\cal C}_{0s1/2}^2$. 
This shows that the reaction is purely peripheral, since it demonstrates that the breakup does not probe the short-range physics of the projectile.

Because the cross sections scale with the ${\cal C}_{0s1/2}^2$ and because the cross section exhibits little sensitivity to the choice of the nuclear part of the optical potential (see \Sec{optpot}), we can infer an ANC by fitting the calculations to the data.
To avoid the regions where the nuclear interaction plays a role and where the $d$ waves, which are not well constrained, might affect the calculation, we focus on the forward-angle region---viz. $\theta<2^\circ$---of the angular distribution, which is restricted to small $^{18}$C-n relative energies---viz. $0\le E\le 0.5$~MeV.
As seen in \Fig{f2}, that region is dominated by the $p$-wave contribution.

In order to extract a reliable value for ${\cal C}_{0s1/2}$ from data it is necessary  to account for the experimental resolution.
This is done by folding the theoretical cross sections with the resolution provided in the experimental paper \cite{Nakamura:1999rp}.
After folding, we scale the calculations to the data.
Minimizing the $\chi^2$ with respect to the scaling factor enables us to infer the ANC:
\begin{equation}
{\cal C}_{0s1/2}=0.81\pm0.02~{\rm fm}^{-1/2}.\label{e2}
\end{equation}
This value, and its uncertainty, are independent of the value of $\sigma$ chosen for the $^{18}$C-n potential \eq{e1}, confirming the independence of the calculations to the short-range physics, and hence the accuracy of the method.

\section{(In)Sensitivity of the calculations to the nuclear optical potentials}\label{optpot}

To test the sensitivity of the calculations to the choice of optical potentials, they have been repeated with different interactions found in the literature.
The results of the previous section followed Typel and Shyam in Ref.~\cite{TS01} and chose for the $^{18}$C-${}^{208}$Pb interaction a potential developed by Buenerd \etal\ to reproduce the elastic scattering of $^{13}$C off $^{208}$Pb at 390~MeV ($30A$ MeV).
The fact that this is a rather different energy than the one employed in Refs.~\cite{Nakamura:1999rp,Nakamura:2003cyk} is ignored, but the radii are scaled to the actual size of the core of the projectile.
For the n-${}^{208}$Pb interaction, in the previous section we also followed Typel and Shyam and use the Becchetti and Greenlees global optical potential (BG) \cite{BG69}.

For a second $^{18}$C-${}^{208}$Pb optical potential, we use the one considered by Typel and Shyam in Ref.~\cite{TS01} to simulate the interaction between $^{11}$Be and ${}^{208}$Pb at $70A$~MeV. That potential is based on an $\alpha$-${}^{208}$Pb potential developed by Bonin \etal\ to reproduce that elastic scattering at 288~MeV ($72A$ MeV), from which we rescale the radii to account for the size of the nucleus.
As a second n-${}^{208}$Pb potential choice, we opt for the Koning-Delaroche global optical potential (KD) \cite{KD03}.

\begin{figure}[h]%
\centering
\includegraphics[width=0.49\textwidth]{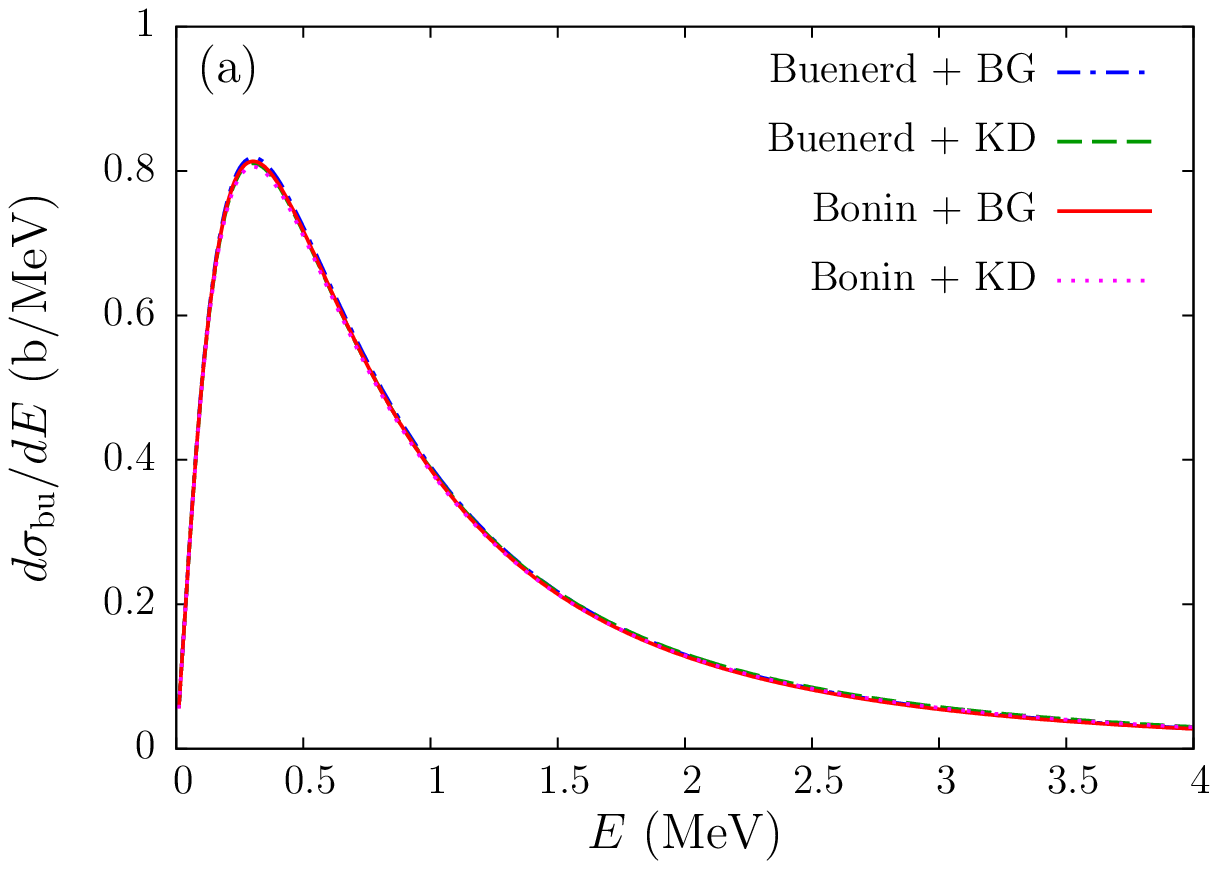}
\includegraphics[width=0.49\textwidth]{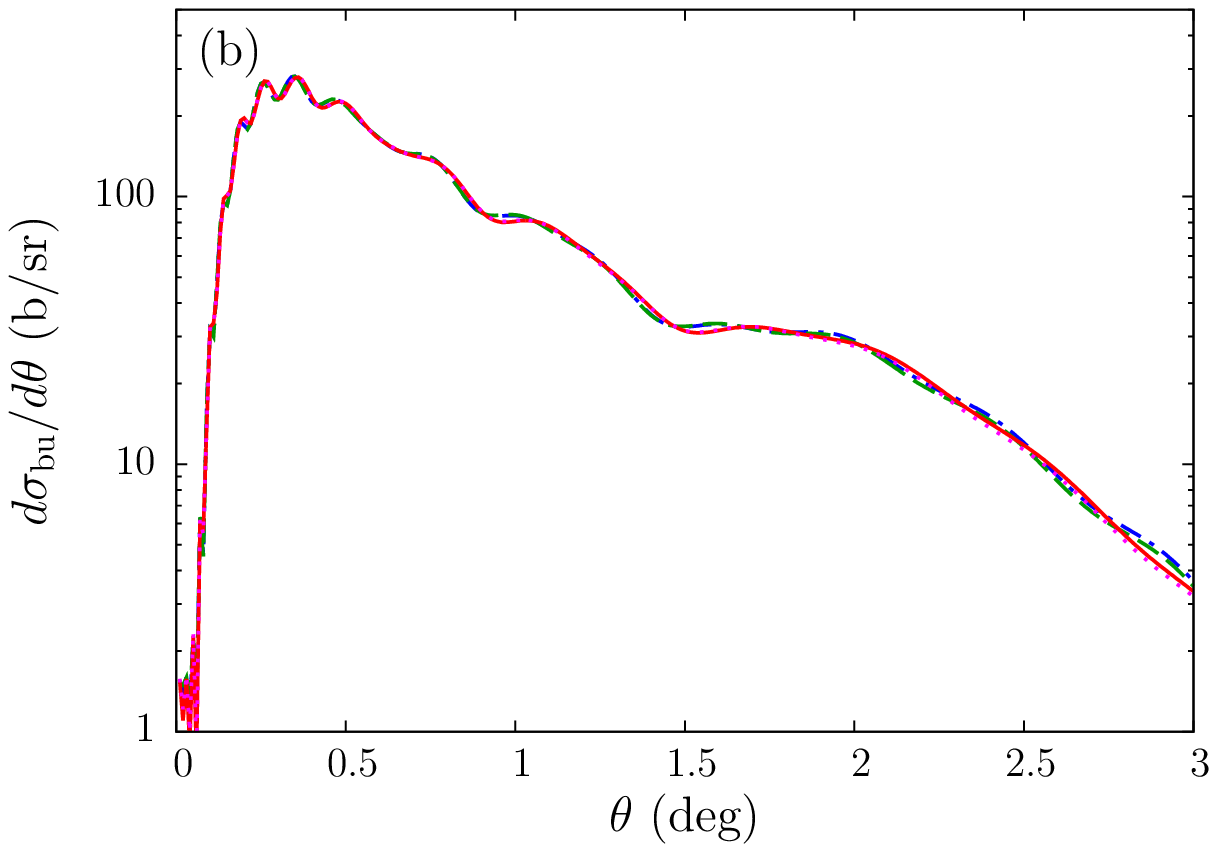}
\caption{Influence of the optical potential choice on the breakup cross section of $^{19}$C on ${}^{208}$Pb at $67A$ MeV;
(a) energy distribution; (b) angular distribution.}\label{f3}
\end{figure}

The results of these different calculations obtained with the LO $^{18}$C-n potential with $\sigma=1.5$~fm are shown in \Fig{f3}.
It is clear that these choices have very limited influence on this Coulomb-dominated reaction.
We note that if we strictly follow Typel and Shyam's procedure from Ref.~\cite{TS01} and do not adjust the radius of the projectile carbon nucleus then we obtain a higher cross section than 
is seen here. The size of the core is an important parameter in these calculations, even if the results are not sensitive to the 
functional form of the potential's radial dependence.

\section{NLO calculations}\label{NLO}

\begin{figure}[h]%
\centering
\includegraphics[width=0.49\textwidth]{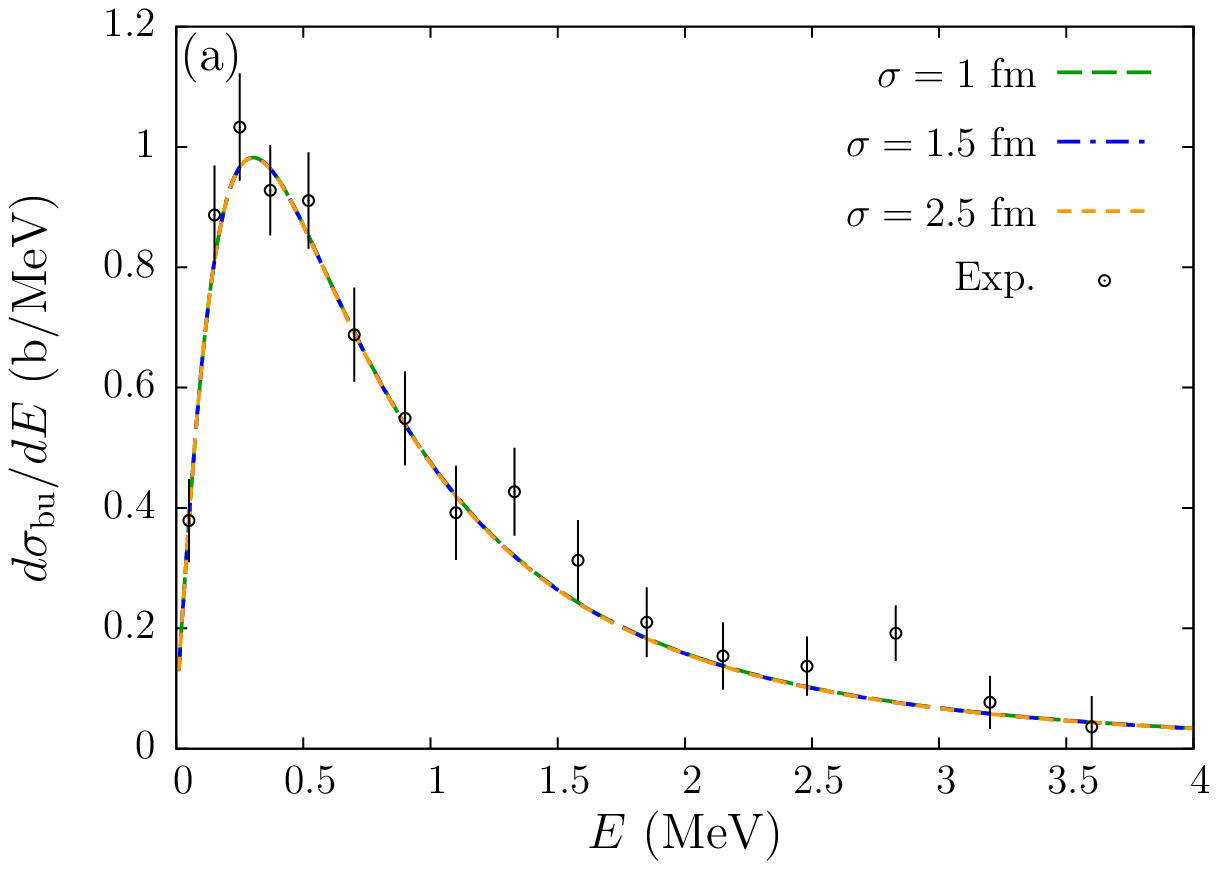}
\includegraphics[width=0.49\textwidth]{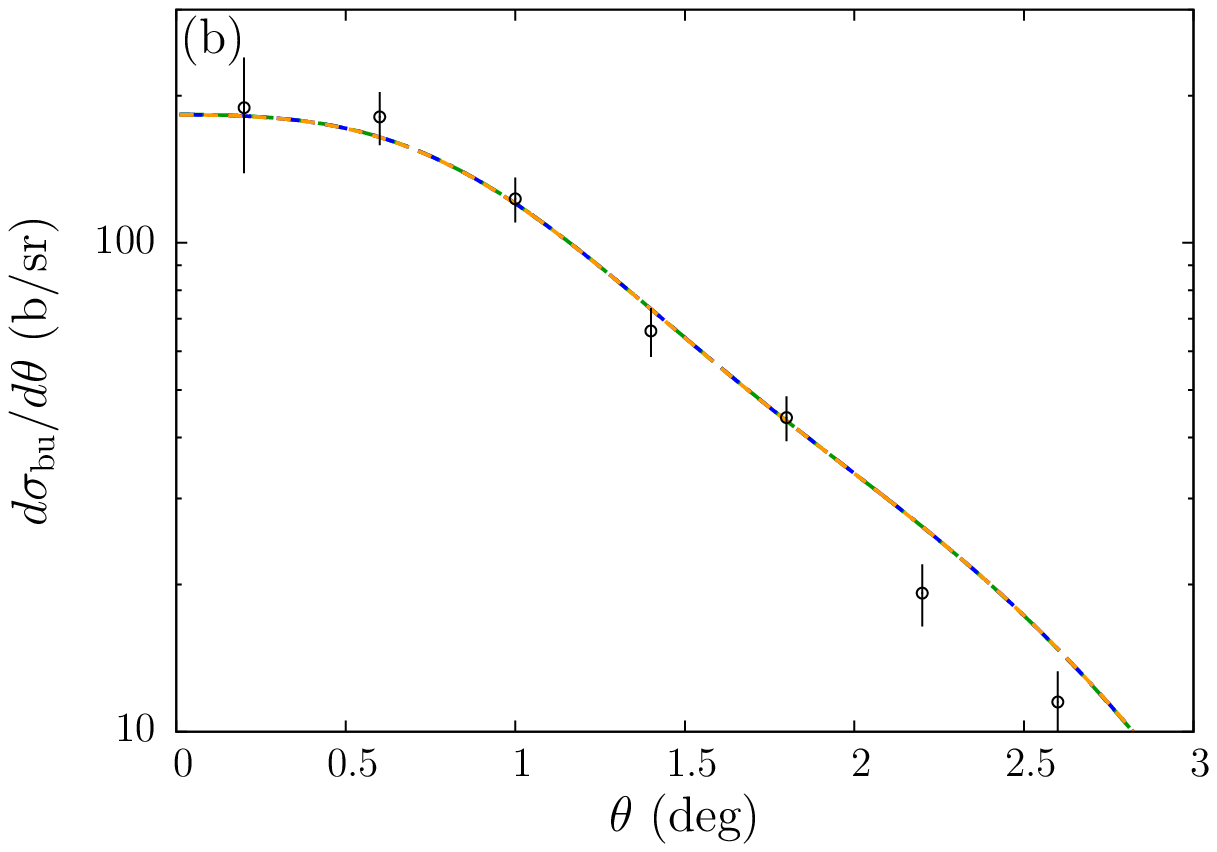}
\caption{NLO calculations of the breakup of $^{19}$C on ${}^{208}$Pb at $67A$ MeV compared to the data of Ref.~\cite{Nakamura:1999rp}.
NLO Halo-EFT $^{18}$C-n potentials are fitted to reproduce the binding energy and the ANC inferred from the comparison of the LO calculations to the data (angular distribution restricted to forward angles);
(a) energy distribution and (b) angular distribution.
In both cases the calculations have been folded by the experimental resolution \cite{Nakamura:1999rp}.}\label{f4}
\end{figure}

At NLO the ${}^{18}$C-neutron potential takes the form:

\begin{equation}
V_{\rm NLO}(r;\sigma)=\frac{1}{(2 \pi \sigma^2)^{3/2}}\left[ \tilde{C_0}(\sigma) \exp\left(-\frac{r^2}{2 \sigma^2}\right) + C_2(\sigma) r^2  \exp\left(-\frac{r^2}{2 \sigma^2}\right)\right],
\label{eq:NLOV}
\end{equation}
where the parameter $\tilde{C_0}$ is not necessarily---indeed not usually---the same as the parameter $C_0$. This time we consider potentials with different $\sigma$s and, in each case adjust them to produce $S_{\rm n}=0.58$ MeV and ${\cal C}_{0s1/2}=0.81~{\rm fm}^{-1/2}$. We achieve this for $\sigma=1.0$, $1.5$, and $2.5$ fm. 
The resulting cross sections predicted by the CCE are now  completely independent of $\sigma$, see \Fig{f4}, where the theoretical cross sections have been folded with the experimental resolution \cite{Nakamura:1999rp}.
Moreover, despite the fact that we fit the ANC only to the forward-angle region of the angular distribution limited to $E\le 0.5$~MeV,  we find that all calculations match the experimental energy distribution over nearly the entire experimental energy range, viz. out to $E=4$~MeV.
The excellent agreement with experiment, and the insensitivity of the NLO results to the regulator $\sigma$, confirm the value of the ANC we inferred from the data using our LO calculations.
This also shows that it is not necessary to go beyond NLO to explain the main features of the data.

Note that no NLO potential could be found for $\sigma=0.5$~fm, i.e, we could not find parameters to fit simultaneously the binding energy and the ANC inferred from the data.
This is a realisation of the Wigner bound~\cite{Wigner:1955zz,Phillips:1996ae,HL10} for this system: for any ${\cal C}_{0s1/2}^2$ larger than $2 (2 \mu S_{\rm n}/\hbar^2)^{1/2}$
the integral of the asymptotic wave function from zero to infinity is larger than one. It follows that for small enough $\sigma$ it is simply impossible to produce a normalisable wave function with this ANC$^2$.

Significant discrepancies between theory and experiment appear at about $E\approx1.3$~MeV and 2.8~MeV in the energy distribution.
At those energies the data seem to be notably larger than our calculations.
The large error bars on the experimental data leave open the possibility that these are statistical fluctuations and not due to an effect of final-state interactions~\footnote{Final-state interactions in the $s$ wave are actually accounted for in our CCE calculation, but are a small effect.}. 
However these deviations could hint at the presence of resonances in the ${}^{19}$C system at these energies. 
Refs.~\cite{Sat08,Hwa17} suggest the existence of a $\fial^+$ resonance at either $E=1.42(10)$~MeV \cite{Hwa17} or $E=1.46(10)$~MeV \cite{Sat08}.
This state might have a dominant single-particle structure with a $^{18}$C core in its $0^+$ ground state and neutron in a $d_{5/2}$ resonance and could  significantly affect the breakup cross section \cite{Fuk04,CGB04,CPH22}.
Within the usual Halo-EFT power counting, it would therefore enter beyond NLO.
Following what has been done in Refs.~\cite{CPH18,CPH22} an extension of this work could study this possibility.

At large angles in the angular distribution, we also observe that the calculations slightly overestimate the data.
This is a region where the nuclear interaction plays a more significant role, see \Fig{f3}(b), and hence is subject to caution because this difference might be related to the choice of optical potentials.
It could also come from the $^{18}$C-n final-state interaction in the $d$ wave, which is not constrained at NLO; see \Fig{f2}(b).

The good agreement with experiment suggests that, in absence of more precise measurements, a Halo-EFT description at NLO is both necessary and sufficient to describe most of the breakup data.

\section{Sensitivity to the binding energy}\label{BE}

The binding energy quoted in the most recent atomic mass database \cite{Masses212} exhibits a rather large uncertainty: $S_{\rm n}=0.58 \pm 0.09$ MeV. 
To gauge the influence of this observable on the calculations, we repeat breakup calculations using LO Halo-EFT $^{18}$C-n potentials fitted to the lower ($S_{\rm n}=0.49$~MeV) and upper ($S_{\rm n}=0.67$~MeV) end of this 68\% confidence interval.
We consider $\sigma=1.5$ fm for this test.

\begin{figure}[h]%
\centering
\includegraphics[width=0.49\textwidth]{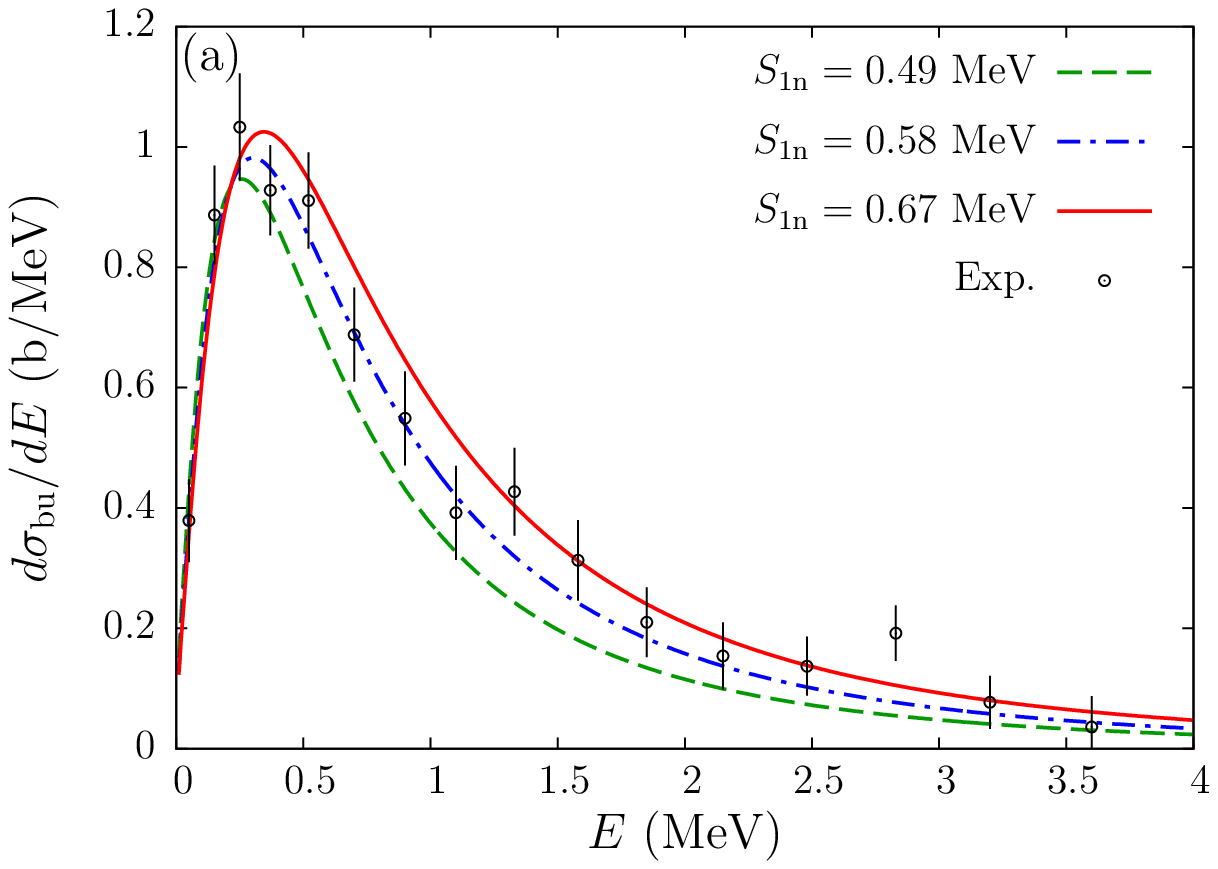}
\includegraphics[width=0.49\textwidth]{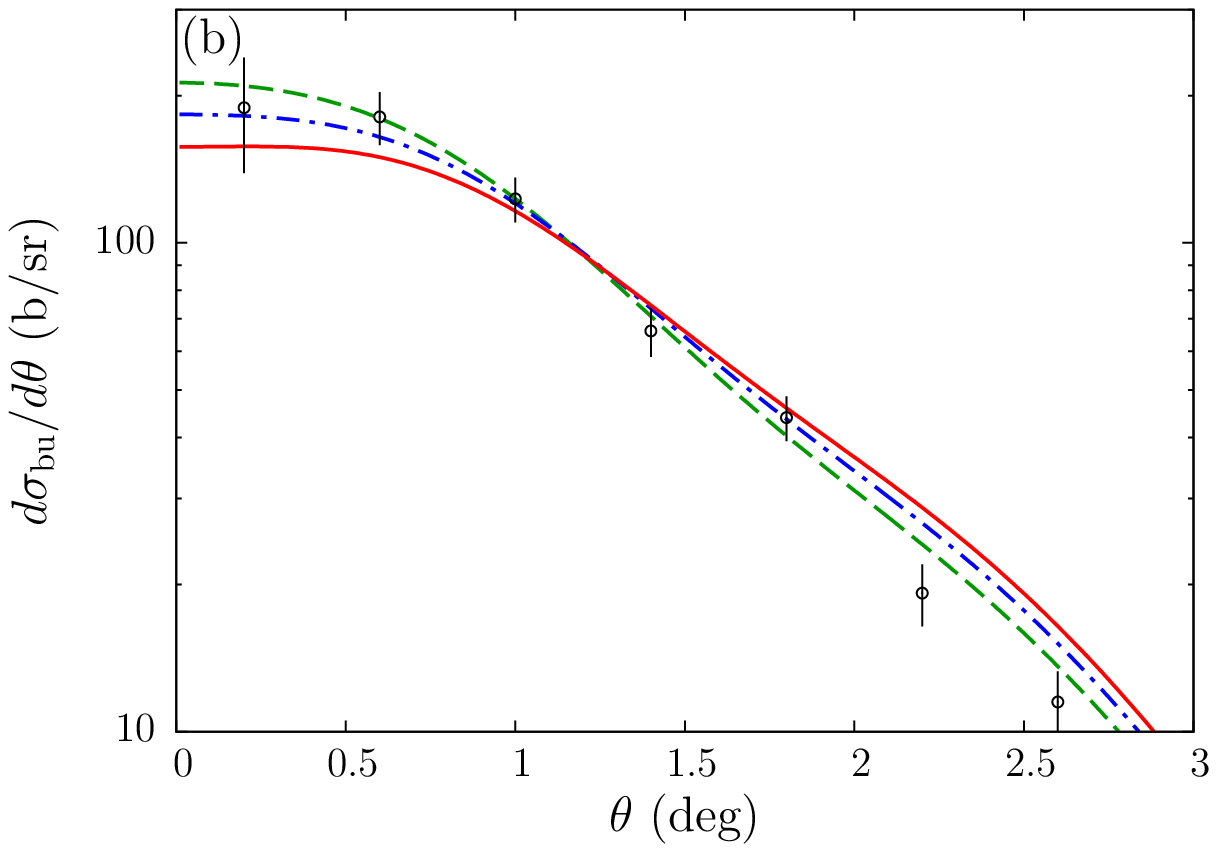}
\caption{Sensitivity of breakup calculation to the $^{18}$C-n binding energy for $^{19}$C impinging on ${}^{208}$Pb at $67A$ MeV.
LO Halo-EFT $^{18}$C-n potentials are fitted to reproduce three binding energies at the centre (0.58~MeV, blue dash-dotted line), lower bound (0.49~MeV, green dashed line) and higher  bound (0.67~MeV, red solid line) of the experimental uncertainty range \cite{Masses212}.
The calculations have been scaled to the data as explained in \Sec{NLO};
(a) energy distribution; (b) angular distribution.
In both cases the calculations have been folded with the experimental resolution \cite{Nakamura:1999rp}.}\label{f5}
\end{figure}

\Fig{f5} displays the results folded with the experimental resolution and fitted to the data by rescaling the CCE calculation.
The ANCs hence obtained differ significantly from the one quoted above:
${\cal C}_{0s1/2}(S_{\rm n}=0.49\,{\rm MeV})=0.62\pm0.02$~fm$^{-1/2}$ and 
${\cal C}_{0s1/2}(S_{\rm n}=0.67\,{\rm MeV})=1.02\pm0.03$~fm$^{-1/2}$.
This shows that the ANC and binding energy are strongly correlated, as one would expect from the LO Halo EFT relation \cite{SCB10,Hammer:2017tjm}
\begin{equation}
{\cal C}_{0s1/2}^2(S_{\rm n}) = 2 \sqrt{\frac{2 \mu S_{\rm n}}{\hbar^2}}\left[1 +{\cal O}\left(\sqrt{\frac{2 \mu S_{\rm n}}{\hbar^2}} \sigma\right) \right]
\label{eq:ANCscaling}
\end{equation}
We note that, while the strict scaling fo the ANC-squared at LO is with $\sqrt{S_{\rm n}}$, the higher-order terms indicated in Eq.~(\ref{eq:ANCscaling}) are ultimately quite important in the case of ${}^{19}$C. The ANC-squareds inferred for different binding energies scale markedly more strongly with $S_{\rm n}$ than $\sqrt{S_{\rm n}}$.

Although the prediction with the lowest binding energy seems to better reproduce the angular distribution throughout the entire experimental angular range [see the green dashed line in \Fig{f5}(b)], the corresponding energy distribution does not fit the data at $E>0.5$~MeV [see \Fig{f5}(a)].
Using a higher binding energy leads to less good agreement with the data in both observables (red solid lines in \Fig{f5}).
This suggests that the actual binding energy is probably close to the central value we have considered up to \Sec{NLO}, viz. $S_{\rm n}=0.58$~MeV.

It also indicates that, with more thorough uncertainty quantification, including an estimate of the impact of higher-order effects in the EFT~\cite{FPW15,FKP15,SNP22}, and the uncertainty due to the choice of optical potential, analysis of these data could yield a new, more precise, value, for the ${}^{19}$C one-neutron separation energy.

\section{Conclusion and Needed Future Work}\label{Conclusion}

Many experiments have shown that $^{19}$C exhibits a clear one-neutron halo structure in its $\half^+$ ground state \cite{Baz95,Mar96,BBB98,Nakamura:1999rp,Nakamura:2003cyk,Sat08,Hwa17}.
However, in contrast to the well-studied cases of $^{11}$Be and $^{15}$C, there is still much to learn about this nucleus, including its one-neutron separation energy $S_{\rm n}$.
In this paper, we present a new analysis of the Coulomb breakup of $^{19}$C on ${}^{208}$Pb at $67A$ MeV, which has been measured at RIKEN \cite{Nakamura:1999rp}.
To this aim, we have used a Halo-EFT description of the projectile within the Coulomb Corrected Eikonal approximation (CCE), which has shown to provide reliable cross sections for this kind of reaction \cite{CBS08}, while exhibiting a small numerical cost.

As expected these cross sections are strongly dominated by an E1 transition from the $0s_{1/2}$ ground state of the nucleus towards its $^{18}$C-n continuum.
Being Coulomb dominated, they exhibit a minor dependence to the optical potentials used to simulate the nuclear interaction between the projectile constituents ($^{18}$C and n) and the ${}^{208}$Pb target.

Using a LO description of $^{19}$C, we have found out that the calculated cross sections are nearly proportional to the square of the ANC of the radial $^{18}$C-n wave function ${\cal C}_{0s1/2}$.
This clearly shows that the reaction is purely peripheral in the sense that it probes only the tail of the ground state wave function.
It also indicates that an ANC for the actual nucleus can be inferred from the data.
To reduce the uncertainty related to the choice of the optical potentials as well as to avoid the influence of the $d$-wave continuum, we select forward-angle data at low $^{18}$C-n energy to scale our calculations to the experiment.
The value of the ANC hence obtained, ${\cal C}_{0s1/2}=0.81\pm0.02$~fm$^{-1/2}$, is independent of the Halo-EFT regulator $\sigma$.
NLO descriptions of $^{19}$C fitted to reproduce both $S_{\rm n}$ and that value of ${\cal C}_{0s1/2}$ provide an excellent agreement with the data on nearly their entire energy and angular ranges, independently of the value of $\sigma$.

Additional tests have shown a strong dependence of the calculations to the binding energy of the nucleus, which, unfortunately is not well known experimentally.
However, our tests show that a systematic analysis of Coulomb-breakup data, e.g., through Bayesian methods \cite{FPW15,FKP15,SNP22}, could provide a significant constraint on that structure observable.
Thanks to its small computational cost and accurate description of the reaction process, the CCE would be the ideal reaction-dynamics treatment for such a future statistical analysis.

This theoretical study extends a series of analyses of reactions involving one-neutron halo nuclei, in which a Halo-EFT description of the exotic nucleus is coupled to realistic models of reactions \cite{CPH18,YC18,MYC19,HC21}.
Our work confirms the validity of this approach for the Coulomb breakup of $^{19}$C, and shows that crucial nuclear-structure information can be inferred from such a study.
Unfortunately, the experimental uncertainty of the RIKEN data considered in this work \cite{Nakamura:1999rp} is too large to draw reliable conclusions on these structure observables.
Accordingly, we advocate for new experiments with smaller uncertainties to pin down these values.
Similar breakup data would help us constrain both the binding energy of $^{19}$C and its ANC.
Breakup data on a light target, viz. $^{12}$C or $^9$Be, could help investigate the possible presence of single-neutron resonances in the continuum.
Transfer measurements, such as $^{18}$C(d,p) in inverse kinematics could help constrain the ANC of the ground state, especially if they are measured at low beam energy and forward angles \cite{YC18}.
Knockout measurements with improved uncertainty compared to existing data \cite{Baz95,Mar96,BBB98} would also improve our understanding of this exotic nucleus \cite{HC21}.

\backmatter

\bmhead{Acknowledgments}
This project has initiated as an exercise during the TALENT (Training in Advanced Low Energy Nuclear Theory) summer school on ``Effective Field Theory in Light Nuclei: from Structure to Reactions'', which took place at the Johannes Gutenberg Universit\"at Mainz (JGU, Germany) from 25 July to 12 August 2022 \cite{TALENT22}.
The interesting results obtained by the students led PC to complete their initial calculations and DP and PC to compile the results in this manuscript.
The school has been funded by the Mainz Institute for Theoretical Physics (MITP) and the International Office of the JGU, which we thank for their support during the school.
We also thank the teaching assistants Thomas Richardson and Martin Sch\"afer for their help during the exercise sessions.

In addition, this project has received funding from the Deutsche Forschungsgemeinschaft within the Collaborative Research Center SFB 1245 (Projektnummer 279384907)
and the PRISMA (Precision Physics, Fundamental Interactions and Structure of Matter) Cluster of Excellence.
It has also received support from the US Department of Energy (contract DE-FG02-93ER40756).
N.\ Y.\ was supported in part by the National Science Foundation
under Grant No.\ PHY--2044632.
P.\ C.\ acknowledges the support of the State of Rhineland-Palatinate.


\end{document}